\documentclass[11pt,a4paper]{article}

\usepackage[margin=1in]{geometry}
\usepackage{times}
\usepackage{amsmath,amssymb}
\usepackage{graphicx}
\usepackage{booktabs}
\usepackage{hyperref}
\usepackage{url}
\usepackage{xcolor}
\usepackage{setspace}
\usepackage{natbib}
\usepackage{orcidlink}
\hypersetup{
  colorlinks=true,
  linkcolor=blue!50!black,
  citecolor=blue!50!black,
  urlcolor=blue!60!black
}

\title{Open versioned aggregates of the FAA National Wildlife Strike Database (1990--2026):\\
conditional damage rates by animal family, airframe, and airport}

\author{
  Denys Kolomiiets\,\orcidlink{0000-0003-3212-3251}\\
  \small FlightFinder / independent researcher\\
  \small \texttt{denys@himaxym.com} \quad \url{https://himaxym.com}
}

\date{August 2026}

\begin{document}
\maketitle

\begin{abstract}
\noindent\textbf{Background.}
Aircraft--wildlife collisions are a documented aviation hazard.
The U.S.\ FAA National Wildlife Strike Database (NWSD) is the primary public source of U.S.\ reports, but raw extracts are large and rarely released as versioned research aggregates.

\medskip
\noindent\textbf{Methods.}
We convert the public FAA Access export (\texttt{STRIKE\_REPORTS}) to SQLite, define a report as \emph{damaging} when \texttt{DAMAGE\_LEVEL}$\in\{$M, M?, S, D$\}$, fold rows into open aggregates (year, airport, species, aircraft, animal family), and map hierarchical \texttt{SPECIES\_ID} prefixes to animal families.
We analyze all civil reports and a commercial subset (\texttt{AC\_MASS}$\in\{3,4,5\}$).
Rates are per reported strike (not per flight hour), so every count and rate here is conditional on a report having been filed.
Trend statistics use complete incident years 1990--2025; cross-sectional facets cover the full seed, incident years 1990--2026.

\medskip
\noindent\textbf{Results.}
\emph{Numbers are locked to the \texttt{2026-07-04} data freeze (Section~\ref{sec:availability}).}
Reported strikes in 2025 reach a series high---the largest annual total and damaging count in the 1990--2025 record under our definitions (\textbf{24{,}458} total; \textbf{901} damaging).
Species that dominate strike counts differ systematically from those with the highest damage rates: perching songbirds account for 41.1\% of identified strikes but damage aircraft in 1.5\% of reports (95\% CI 1.4--1.6), whereas deer account for 0.7\% of identified strikes yet damage aircraft in 81.8\% of reports (95\% CI 79.8--83.7).
Airport rankings by total strikes (led by Denver International) diverge from rankings by damaging strikes (led by Sacramento International); across the 452 airports with at least 50 reports, reported strike volume and conditional damage rate are negatively rank-correlated (Spearman $\rho=-0.44$, $p<0.001$); large stations plausibly capture minor events more completely, so this is not a hazard ranking.
The share of reports labeled damaging declines over decades, consistent with improved reporting of minor events, while absolute damaging counts still peak in 2025.

\medskip
\noindent\textbf{Conclusions.}
Versioned open aggregates and a documented damage-rate analysis enable independent reuse.
We release aggregates under CC~BY~4.0.

\medskip
\noindent\textbf{Keywords:}
wildlife strike, bird strike, aviation safety, FAA NWSD, open data, damage rate
\end{abstract}

\section{Introduction}
\label{sec:intro}

Aircraft collisions with birds and other wildlife (``wildlife strikes'') can damage airframes, force aborted takeoffs or returns, delay passengers, and---rarely---kill people.
In the United States the FAA National Wildlife Strike Database (NWSD) collects those reports, mixes voluntary and mandatory channels, and publishes annual summaries~\citep{faa_wildlife_1990_2024}.

The operational cost is real even when the event is not fatal.
A damaging report often means an aborted takeoff, a precautionary return, engine inspection, or unscheduled maintenance; repairs, downtime, and delay costs for U.S.\ civil aviation have been estimated in the hundreds of millions of dollars a year~\citep{allan2000cost,faa_wildlife_1990_2024}.
Fatal outcomes are rare relative to how many reports enter the database, but they are not theoretical: over the jet era, bird and other wildlife strikes have been linked to hundreds of deaths and numerous hull losses worldwide~\citep{thorpe2012progress}.
The 2009 ditching of US Airways Flight 1549 in the Hudson River, after a multiple Canada goose ingestion shortly after takeoff, is still the public reminder that even well-managed operations can be hit hard at low altitude~\citep{marra_et_al_2009_miracle}.

Most of the reported activity sits where mitigation is actually feasible.
Reports cluster at low altitude near airports---takeoff, approach, and landing---so airfield land use and habitat management are the main levers operators still control~\citep{dolbeer2006height,blackwell_et_al_2009}.
That does not mean ``more reports'' is the same problem as ``more damage.''
The species that show up most often in the database are not the ones most likely to leave a damaging report: conditional damage varies sharply across taxa with body mass and flocking behaviour~\citep{devault_et_al_2011}.
DeVault and colleagues established that volume--damage split on earlier taxonomic subsets.
What was missing for day-to-day reuse was a national, multi-decade re-measurement with fixed definitions, interval estimates where they help, and tables anyone can recompute without starting from the raw Access dump again.
That is the job of this paper: not a claim that the ecological pattern is new, but a transparent re-run at full NWSD scale and an open release of the aggregates that make the same check cheap for the next person.
Our own annual totals and damaging counts are checked against the official FAA series (Section~\ref{sec:validation}).

Despite how much the community already relies on NWSD~\citep{faa_wildlife_1990_2024,faa_nwsd_portal}, many analyses still begin by re-deriving the same rollups from bulky exports.
Having built the ingest for FlightFinder---zip from wildlife.faa.gov, Access table out, SQLite fold, seed JSON---I kept hitting the same friction: journalists and researchers want year / airport / species tables and a clear \emph{damage-rate} cut, not another copy of the microdata.
We contribute:
\begin{enumerate}
  \item Machine-readable, multi-decade aggregates indexed by year, airport, species, and aircraft model string;
  \item An explicit conditional \emph{damage-rate} view structured by animal family prefixes and aircraft model strings, complete with confidence intervals;
  \item A transparent processing pipeline with fixed damage thresholds and commercial flight filtering;
  \item A citable open release (Zenodo) under CC~BY~4.0.
\end{enumerate}

Two definition choices are worth flagging up front because they shape every rate in the tables.
A report counts as damaging when \texttt{DAMAGE\_LEVEL} is M, M?, S, or D---including uncertain minor damage---so our rates sit a little higher than analyses that drop M?.
Aircraft model rates group on the free-text \texttt{AIRCRAFT} field as the FAA stores it, not on a cleaned ICAO type designator; that matches the public export but mixes spelling variants.

All rates below are shares among \emph{reports that exist}.
We do \emph{not} claim a complete census of every wildlife strike, flight-hour rates, or a ranking of ``most dangerous airports.''

\section{Related work}
\label{sec:related}

\subsection{Official NWSD reporting}
The FAA, in partnership with USDA Wildlife Services, publishes multi-year summaries of wildlife strikes to civil aircraft~\citep{faa_wildlife_1990_2024,faa_wildlife_1990_2022,faa_wildlife_1990_2020}.
These reports are the authoritative narrative of national trends, economic impact estimates, and management recommendations.
The public portal and Access export make microdata available~\citep{faa_nwsd_portal}, but researchers must re-implement cleaning and aggregation for each study.

\subsection{Scientific analyses of wildlife hazards}
A large literature links wildlife ecology to aviation risk.
Dolbeer and collaborators have documented altitude distributions of bird strikes~\citep{dolbeer2006height}, reporting completeness under largely voluntary systems~\citep{dolbeer2015reporting}, and species identification in high-profile accidents~\citep{marra_et_al_2009_miracle}.
Airport land-use and interspecific hazard variation have been treated as planning and management problems~\citep{blackwell_et_al_2009,devault_et_al_2011}.
Broader reviews synthesize global fatality and cost estimates~\citep{sodhi2002birds,thorpe2012progress,allan2000cost,systematic_review_2025}.

\subsection{Open aviation data}
Open flight-tracking (e.g., ADS-B) and multi-source accident databases have improved reproducibility for many aviation-safety questions.
Wildlife-strike research still often depends on one-off scripts against NWSD dumps, with limited DOI-versioned aggregates that separate \emph{strike volume} from \emph{conditional damage rates} by animal family and airframe string.

\subsection{Gap}
We provide a transparent, re-runnable aggregation layer with fixed damage and commercial definitions, family taxonomy tied to FAA \texttt{SPECIES\_ID} prefixes, and open tables suitable for independent reuse---complementing agency PDFs rather than replacing them.

\section{Data sources}
\label{sec:data}

\subsection{Primary source}
We use the U.S.\ FAA National Wildlife Strike Database public export distributed as an Access database at \url{https://wildlife.faa.gov/} (U.S.\ government public domain work).
By construction the NWSD records strikes to \emph{civil} aircraft; military strikes are collected separately by the U.S.\ Department of Defense and are outside this source.
The analysis unit is one row of the \texttt{STRIKE\_REPORTS} table (one reported event).

\subsection{Temporal window}
The frozen seed holds \textbf{347{,}575} reports spanning incident years 1990--2026, of which \textbf{4{,}010} fall in the partial year 2026.
Time-series statistics use complete incident years 1990 through \textbf{2025} inclusive (\textbf{343{,}565} reports); the partial 2026 is excluded from every trend statement.
All cross-sectional facets---airport, species, animal family, and aircraft---are computed over the full seed with the partial 2026 included, because the frozen aggregates store facet totals only and are not re-windowable by year.
The 2026 rows are 1.2\% of the seed and cannot move any reported rate materially, but every cross-sectional $n$ in this paper should be read as covering 1990--2026.
The software snapshot used in this article is recorded in the data availability statement (Section~\ref{sec:availability}).

\subsection{Key fields}
We use (among others) \texttt{INCIDENT\_YEAR}, \texttt{AIRPORT\_ID}, \texttt{AIRPORT}, \texttt{SPECIES}, \texttt{SPECIES\_ID}, \texttt{AIRCRAFT}, \texttt{AC\_MASS}, \texttt{DAMAGE\_LEVEL}, \texttt{PHASE\_OF\_FLIGHT}, \texttt{NR\_FATALITIES}, \texttt{NR\_INJURIES}, and free-text \texttt{REMARKS} for a capped narrative set only.

\subsection{Secondary metadata}
Airport coordinates for maps, when needed, are resolved best-effort via ICAO match against OurAirports (CC0)~\citep{ourairports}.

\section{Methods}
\label{sec:methods}

Implementation details are documented against the FlightFinder open pipeline; core definitions follow.

\subsection{Ingest}
We download the FAA zip archive, export \texttt{STRIKE\_REPORTS} with date-normalized fields, load into SQLite, and compute aggregates offline (no full microdata release except a capped set of severe events with public remarks).
The aggregation script (\texttt{wildlife-aggregate.js}) is provided in the code annex~\citep{code_annex}.

\subsection{Damage definition}
A report is labeled \emph{damaging} if and only if
\[
\texttt{DAMAGE\_LEVEL} \in \{\texttt{M},\,\texttt{M?},\,\texttt{S},\,\texttt{D}\},
\]
i.e., minor, uncertain-minor, substantial, or destroyed.
\texttt{N} (none) and missing values are non-damaging.
Note that \texttt{M?} (uncertain) is counted as damaging---a conservative choice that should be stated when comparing to studies that drop uncertain codes.

\subsection{Cleaning filters}
\begin{itemize}
  \item Airports: drop \texttt{UNKN}/\texttt{UNKNOWN}; require $\ge 50$ reports per airport facet.
  \item Species facets: drop names beginning with ``Unknown''; $\ge 50$ reports.
  \item Aircraft facets: drop empty/\texttt{UNKNOWN}; $\ge 50$ reports for storage; damage-rate chart uses $\ge 2{,}000$ reports.
  \item Airport slug collisions (e.g., annotated ICAO variants mapping to the same slug) keep the larger total.
\end{itemize}
The $\ge 50$-report threshold applies to the published species table, not to the animal-family roll-ups: family counts are folded from every identified report, so family strikes sum to 209{,}778 while the 287 species facets that clear the threshold sum to 203{,}407.
Species below the threshold are therefore absent from the species table but still present inside their family.

\subsection{Commercial segment}
Segment \texttt{all} uses every civil report row.
Segment \texttt{commercial} restricts to FAA aircraft mass codes \texttt{AC\_MASS}$\in\{3,4,5\}$, i.e.\ larger transport-category aircraft (regional jets, airliners, and heavy transports).

\subsection{Animal family mapping}
Hierarchical FAA \texttt{SPECIES\_ID} prefixes map to families (deer, bats, other mammals, waterfowl, raptors, owls, gulls/terns, shorebirds, pigeons/doves, songbirds, herons/egrets, other).
Two-character prefixes take precedence over one-character parents.
Only identified species contribute to family rates.
The mapping is defined in \texttt{wildlifeFamilies.js} (code annex~\citep{code_annex}) and reproduced in Appendix~\ref{app:families}.

\subsection{Metrics}
Family damage rate $= 100\times\textit{damaging}/\textit{strikes}$ within family and segment.
Model damage rate $= 100\times\textit{damaging}/\textit{total}$ for aircraft free-text strings with at least 2{,}000 reports.
These are \textbf{per reported strike}, not per flight hour or movement.

\subsection{Uncertainty and rank statistics}
\label{sec:uncertainty}
Damage rates are binomial proportions; we report 95\% confidence intervals using the Wilson score method, which behaves well for the small proportions and large denominators that dominate these data.
To quantify the divergence between strike volume and conditional damage severity, we report Spearman rank correlations $\rho$ between facet volume (total reports) and damage rate, with two-sided $p$-values from the $t$ approximation $t=\rho\sqrt{(n-2)/(1-\rho^2)}$.
All intervals and correlations are computed directly from the frozen aggregate tables by \texttt{compute\_stats.js} (code annex~\citep{code_annex}); they describe reporting-conditional proportions, not exposure-normalized risk (Section~\ref{sec:limitations}).
Three further caveats apply to these statistics.
First, the intervals are nominal and descriptive: the 452 airport, 24 family, and 28 aircraft-model intervals are reported as per-facet uncertainty, not as a pre-registered family of hypothesis tests, and no correction for multiple comparisons is applied---individual facets that look extreme should be read as hypothesis-generating.
Second, we do not report confidence intervals for $\rho$; the frozen aggregates carry facet totals rather than the row-level data a resampling interval would need, so only the point estimate and its $p$-value are given.
Third, the rank correlations are unweighted: an airport with 50 reports enters the airport correlation with the same weight as Denver International with 11{,}634, so the coefficient describes the ordering of facets, not a report-weighted association.

\subsection{Software}
Aggregates are folded into a JSON seed and loaded into SQLite tables \texttt{wildlife\_*} for exploration and figure generation.
The five processing scripts---aggregation (\texttt{wildlife-aggregate.js}), family taxonomy (\texttt{wildlifeFamilies.js}), freeze CSV export (\texttt{export\_freeze.py}), figure generation (\texttt{make\_figures.js}), and interval/correlation statistics (\texttt{compute\_stats.js})---are released as an open code annex~\citep{code_annex}; they operate only on the public FAA source and the derived tables, and the full FlightFinder application is not required to reproduce the aggregates.

\section{Results}
\label{sec:results}

Time-series statistics below cover complete incident years 1990--2025.
Facet statistics---airport, species, animal family, aircraft, and phase of flight---cover the full frozen seed, incident years 1990--2026 (Section~\ref{sec:data}).
All counts and rates are conditional on a report having been filed.

\subsection{National time series}
\begin{figure}[t]
  \centering
  \includegraphics[width=0.92\linewidth]{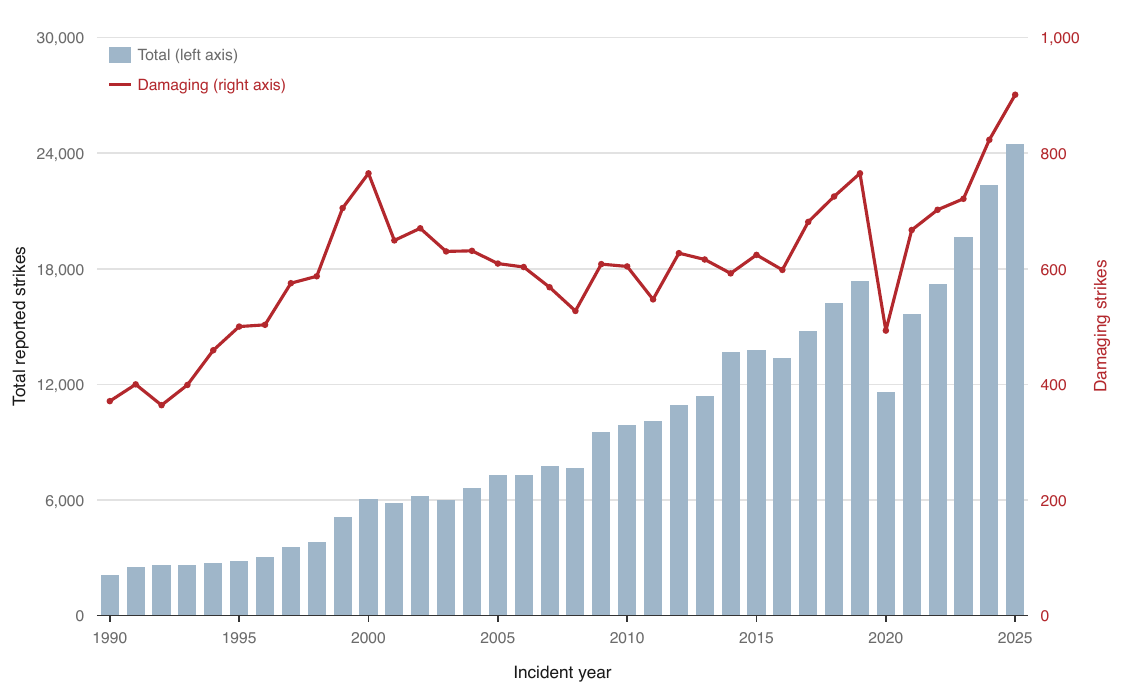}
  \caption{U.S.\ FAA NWSD annual total and damaging reports, 1990--2025 (freeze \texttt{2026-07-04}).}
  \label{fig:yearly}
\end{figure}

Figure~\ref{fig:yearly} shows annual totals and damaging counts for 1990--2025.
Calendar year 2025 records \textbf{24{,}458} reported strikes and \textbf{901} damaging reports.
Across 1990--2025 the database contains \textbf{343{,}565} reports (\textbf{21{,}809} damaging).
The share of reports labeled damaging falls from \textbf{17.5\%} in 1990, the first year of the record, to \textbf{3.7\%} in 2025, consistent with more complete reporting of minor events~\citep{dolbeer2015reporting}, while absolute damaging counts still peak in 2025.
We quote 1990 rather than the series maximum of \textbf{17.7\%} in 1995 so that the comparison runs from the start of the record and not from a chosen high point; on the pooled 1990--1994 baseline the share is 15.8\%, and the direction and rough magnitude of the decline are unchanged.

\subsection{Species volume versus damage}
\begin{figure}[t]
  \centering
  \includegraphics[width=0.78\linewidth]{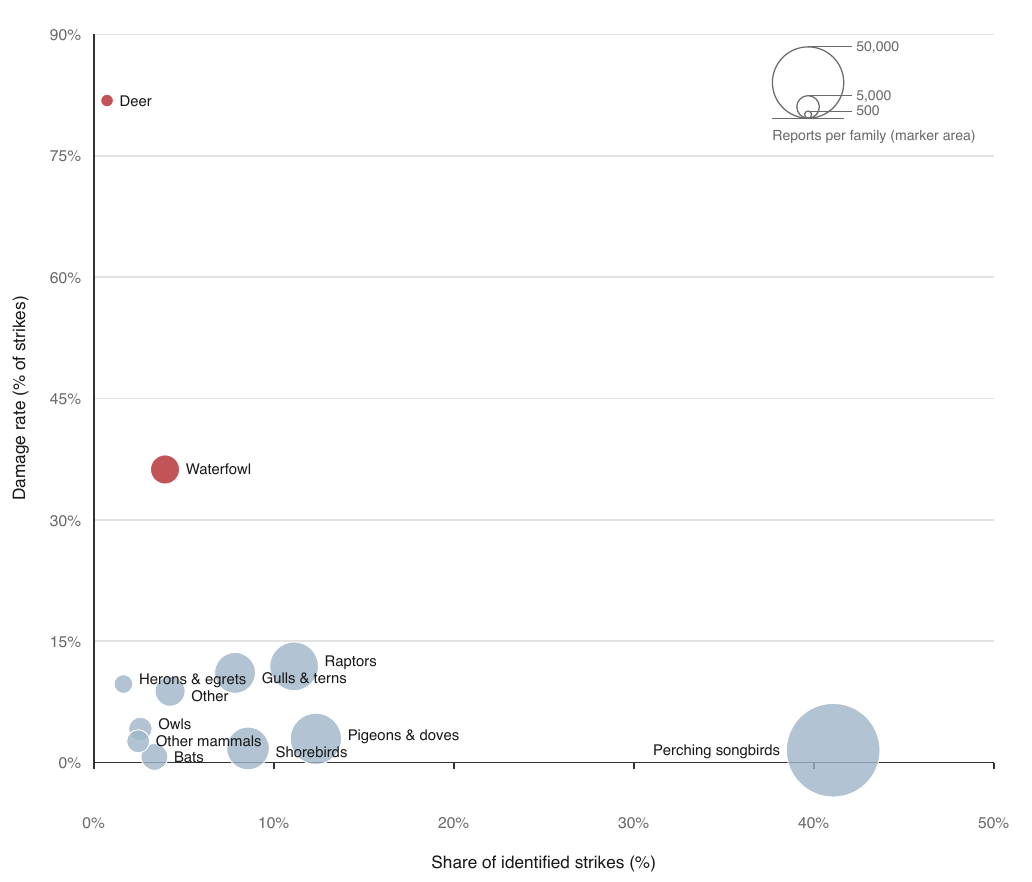}
  \caption{Identified animal families: share of strikes vs damage rate (civil segment; incident years 1990--2026; freeze \texttt{2026-07-04}). Marker radius is proportional to the square root of the family's report count, so marker area is strictly proportional to the number of reports; the nested circles in the upper right are a size legend drawn at the same scale, labeled with reference report counts. Markers are red where the damage rate is at or above 20\% and light steel below it.}
  \label{fig:family}
\end{figure}

Denominators first, so that every share below can be reconstructed.
The frozen seed holds \textbf{347{,}575} reports; \textbf{209{,}778} of them (\textbf{60.4\%}) carry a usable \texttt{SPECIES\_ID} and are folded into an animal family, and the remaining 39.6\% are unidentified.
Family shares in this subsection are shares of those 209{,}778 identified reports; family damage rates are shares of each family's own reports.
Figure~\ref{fig:family} contrasts each animal family's share of identified strikes with its damage rate (all civil segment).
Families that dominate counts are not those with the highest damage rates.
Perching songbirds account for 41.1\% of identified strikes but are damaging in only 1.5\% of reports (95\% CI 1.4--1.6, $n=86{,}172$); deer account for only 0.7\% of identified strikes, yet 81.8\% of reported deer strikes are coded damaging (95\% CI 79.8--83.7, $n=1{,}562$), and waterfowl 36.2\% (95\% CI 35.2--37.3, $n=8{,}312$).
The confidence intervals of the high- and low-severity families are separated by more than an order of magnitude, so the volume--damage divergence is not attributable to sampling noise.
At the family level the rank correlation between share of identified strikes and damage rate is negative but not statistically significant (Spearman $\rho=-0.34$, $n=11$ named families excluding the residual ``Other'' group, $p=0.31$), which is what the divergence claim predicts: the gap is produced by a few high-mass taxa---deer and waterfowl---sitting far above the rest, not by a monotone gradient in which damage falls steadily as reported volume rises.
Eleven families give a rank test almost no power, so the coefficient is reported for completeness rather than as evidence; the family-level claim rests on the separated confidence intervals above.

\begin{figure}[t]
  \centering
  \includegraphics[width=0.78\linewidth]{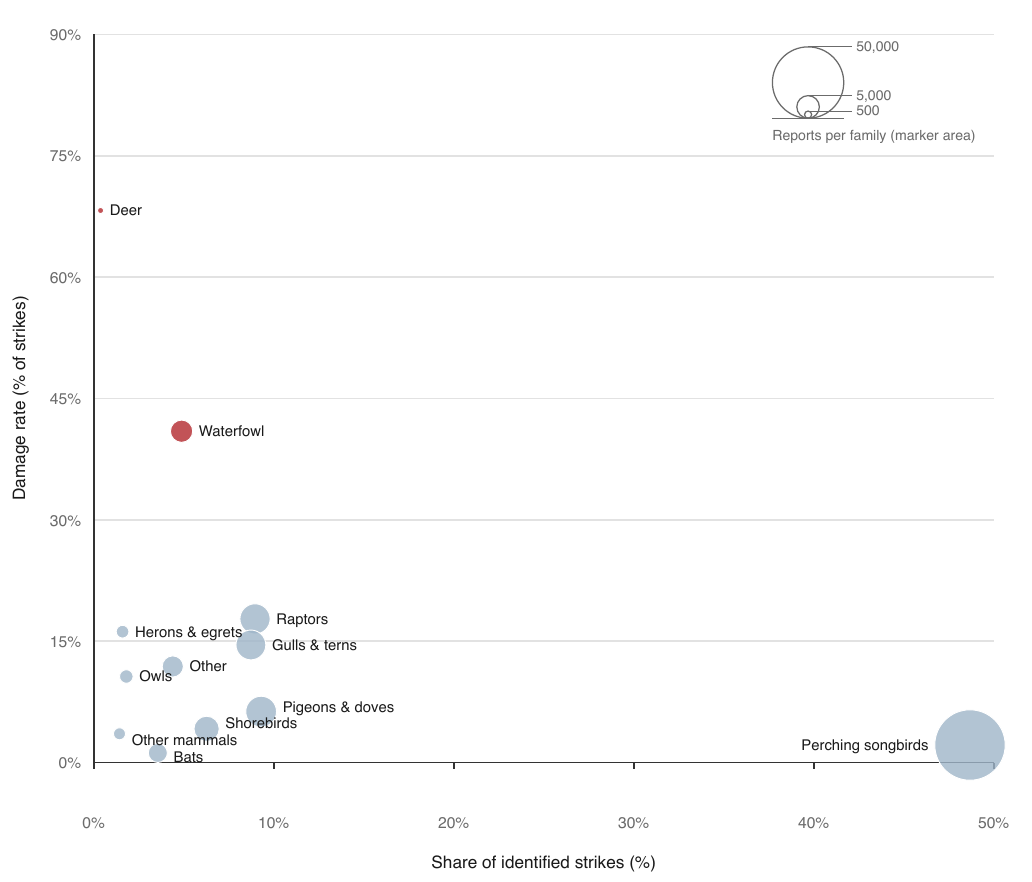}
  \caption{Commercial segment (\texttt{AC\_MASS} 3--5): family strike share vs damage rate (incident years 1990--2026). Marker area is strictly proportional to the number of reports, the upper-right nested circles are a size legend with reference report counts, and red marks a damage rate at or above 20\%, as in Figure~\ref{fig:family}. Both panels are drawn with the same scale factor, so marker areas are directly comparable between the two figures.}
  \label{fig:family-comm}
\end{figure}

The commercial-only segment (mass codes 3--5) is shown in Figure~\ref{fig:family-comm}; the same volume--damage divergence holds, with deer (68.2\%, 95\% CI 63.4--72.7, $n=384$) and waterfowl (41.0\%, 95\% CI 39.6--42.3, $n=4{,}884$) again dominating conditional damage.

\subsection{Aircraft models}
\begin{figure}[t]
  \centering
  \includegraphics[width=0.92\linewidth]{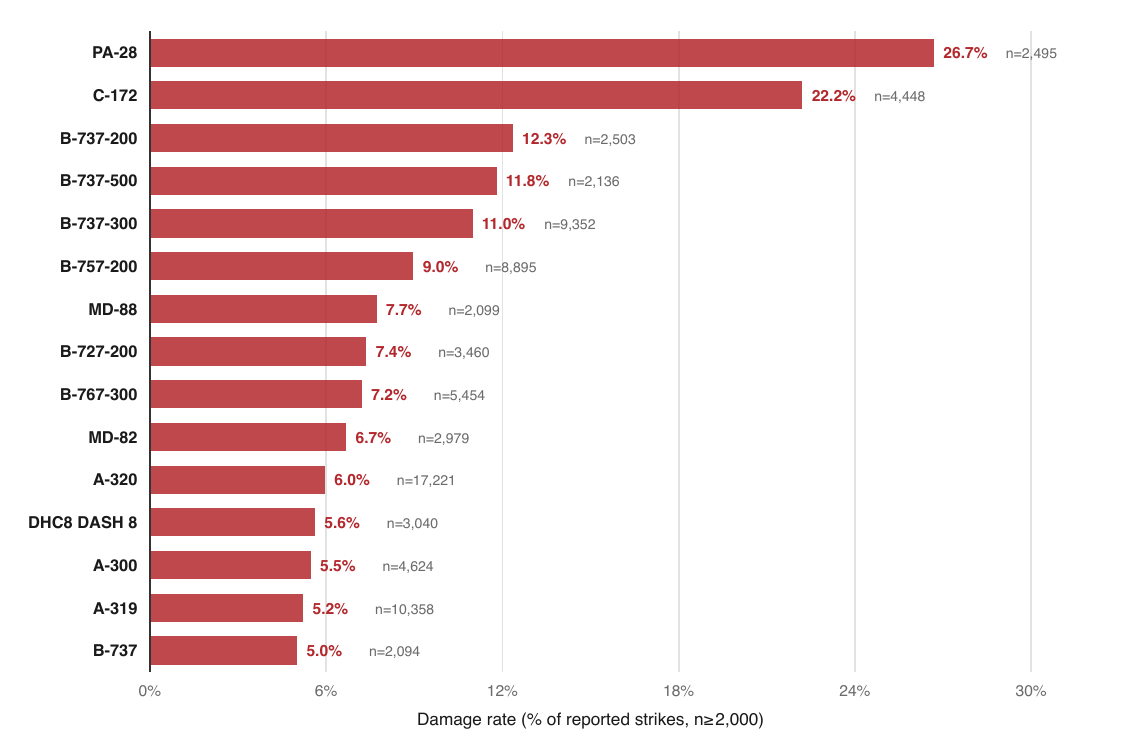}
  \caption{Damage rate by aircraft free-text model string ($n\ge 2000$; incident years 1990--2026; freeze \texttt{2026-07-04}). The panel shows the 15 highest-rate strings of the 28 that meet the threshold; the remaining 13 fall below the lowest bar shown.}
  \label{fig:models}
\end{figure}

For free-text aircraft strings with $\ge 2{,}000$ reported wildlife strikes, we report damage rates (Figure~\ref{fig:models}).
The highest rates fall on light piston aircraft---the Piper PA-28 (26.7\%, 95\% CI 25.0--28.5, $n=2{,}495$) and Cessna~172 (22.2\%, 95\% CI 21.0--23.5, $n=4{,}448$)---while the most frequently reported modern airliners sit far lower despite much larger reported strike counts: the Airbus A320 (6.0\%, 95\% CI 5.6--6.3, $n=17{,}221$), Boeing 737-800 (4.4\%, 95\% CI 4.0--4.7, $n=14{,}917$), 737-700 (4.1\%, 95\% CI 3.7--4.4, $n=12{,}964$), and Embraer E170 (3.7\%, 95\% CI 3.4--4.1, $n=12{,}385$).
Across the 28 model strings meeting the threshold, reported strike volume and damage rate are negatively but not significantly rank-correlated (Spearman $\rho=-0.34$, $n=28$, $p=0.07$); at this sample size the association is a suggestive ordering rather than an established effect.
This ordering is consistent with airframe mass and energy tolerance---a given bird is a larger fraction of a $\sim$1-tonne light aircraft than of a transport jet---but it is confounded: we have no flight-hour or movement denominator, small aircraft may sustain and report damage differently, and older narrow-body variants (e.g.\ the 737-200 and 737-300 at 12.3\% and 11.0\%) mix reporting-era and fleet effects.
These rates are therefore descriptive, not exposure-normalized aircraft risk.

\subsection{Airports}
Table~\ref{tab:airports} lists top airports by total and by damaging counts.
Denver International leads total volume (11{,}634 reports, 253 damaging; damage rate 2.2\%, 95\% CI 1.9--2.5); Sacramento International, only eighth by volume (4{,}334), leads all airports by damaging count (491; damage rate 11.3\%, 95\% CI 10.4--12.3)---illustrating rank divergence.
The divergence is systematic rather than anecdotal: across all 452 airports with at least 50 reports, total reported strike volume and conditional damage rate are negatively rank-correlated (Spearman $\rho=-0.44$, $n=452$, $p<0.001$)---busier airports tend to record a \emph{lower} share of damaging outcomes, consistent with more complete capture of minor events at large reporting stations and with fleet-mix differences.

\begin{table}[t]
  \centering
  \caption{Top~10 U.S.\ airports by total reported wildlife strikes, with damaging counts and damage rate (incident years 1990--2026; freeze \texttt{2026-07-04}). Volume rank diverges sharply from damage: Denver leads volume at a 2.2\% damage rate, whereas Sacramento---only eighth by volume---records the most damaging strikes of any airport (491; 11.3\%).}
  \label{tab:airports}
  \begin{tabular}{rllrrr}
    \toprule
    \# & ICAO & Airport & Total & Damaging & Rate \\
    \midrule
    1 & KDEN & Denver Intl & 11{,}634 & 253 & 2.2\% \\
    2 & KDFW & Dallas/Fort Worth Intl & 8{,}992 & 279 & 3.1\% \\
    3 & KORD & Chicago O'Hare Intl & 7{,}331 & 271 & 3.7\% \\
    4 & KJFK & New York (JFK) & 6{,}705 & 333 & 5.0\% \\
    5 & KMEM & Memphis Intl & 5{,}518 & 200 & 3.6\% \\
    6 & KSLC & Salt Lake City Intl & 4{,}391 & 378 & 8.6\% \\
    7 & KATL & Atlanta (Hartsfield--Jackson) & 4{,}351 & 146 & 3.4\% \\
    8 & KSMF & Sacramento Intl & 4{,}334 & \textbf{491} & \textbf{11.3\%} \\
    9 & KDTW & Detroit Metro & 4{,}322 & 117 & 2.7\% \\
    10 & KMCO & Orlando Intl & 4{,}302 & 317 & 7.4\% \\
    \bottomrule
  \end{tabular}
\end{table}

\subsection{Phase of flight}
Phase-of-flight rollups are computed over the full seed snapshot (347{,}575 rows, including the partial 2026 year), because phase totals are not re-windowed in the frozen aggregates; they are descriptive context rather than trend statistics.
Phase is also frequently missing: the phase categories sum to \textbf{210{,}664} reports, so \textbf{136{,}911} reports (\textbf{39.4\%} of the seed) carry no phase code at all and appear in no phase category.
\textbf{Approach} is the most frequent coded phase, with \textbf{89{,}595} reports---\textbf{42.5\%} of the 210{,}664 reports that carry a phase code, and 25.8\% of all 347{,}575 seed rows.
The first of those two shares is the one to read as ``share of phase-coded reports''; the second must not be read as implying that the other 74.2\% of reports occurred in some other phase, because most of that remainder has no phase recorded at all.
Landing roll follows approach, matching the low-altitude concentration of reported strikes documented for this hazard~\citep{dolbeer2006height}.
Whether phase coding is missing at random with respect to phase is unknown, so the coded-share figure is itself conditional on a phase having been recorded.

\subsection{Validation against the official FAA report}
\label{sec:validation}
Because this pipeline re-implements ingest, cleaning, and aggregation from the raw Access export rather than reusing any FAA tooling, its output can drift from the agency's own tabulations without anyone noticing.
Table~\ref{tab:validation} therefore compares our annual totals and damaging counts, year by year, against the corresponding figures in the most recent official multi-year report, \emph{Wildlife Strikes to Civil Aircraft in the United States, 1990--2024}~\citep{faa_wildlife_1990_2024}, for the ten most recent years that report covers.
The official column is the report's overall annual total, which includes en-route reports and reports with no identified airport, so the two series are counted on the same basis.
Agreement is close: the largest relative deviation over the ten years is 0.398\% (2016), most years differ by only a handful of reports, and damaging counts differ by 0 to 2 reports per year.
The residual differences run in both directions and are of the size expected from late-arriving reports and record revisions accumulating between the extract date behind the FAA report and our 2026-07-04 freeze; they are far too small to affect any claim in this paper.

\begin{table}[t]
  \centering
  \caption{Validation against the official FAA/USDA report \emph{Wildlife Strikes to Civil Aircraft in the United States, 1990--2024}~\citep{faa_wildlife_1990_2024}, for the ten most recent years that report covers. ``Diff.'' is this paper's freeze minus the official figure. Maximum relative deviation is 0.398\% (2016); damaging counts agree to within two reports in every year.}
  \label{tab:validation}
  \begin{tabular}{rrrrrrrr}
    \toprule
    & \multicolumn{4}{c}{Total reports} & \multicolumn{3}{c}{Damaging reports} \\
    \cmidrule(lr){2-5} \cmidrule(lr){6-8}
    Year & FAA & This paper & Diff. & Diff.\ (\%) & FAA & This paper & Diff. \\
    \midrule
    2015 & 13{,}771 & 13{,}777 & $+6$  & $+0.044$ & 623 & 624 & $+1$ \\
    2016 & 13{,}325 & 13{,}378 & $+53$ & $+0.398$ & 598 & 598 & $0$ \\
    2017 & 14{,}771 & 14{,}756 & $-15$ & $-0.102$ & 680 & 681 & $+1$ \\
    2018 & 16{,}205 & 16{,}206 & $+1$  & $+0.006$ & 725 & 725 & $0$ \\
    2019 & 17{,}351 & 17{,}348 & $-3$  & $-0.017$ & 765 & 765 & $0$ \\
    2020 & 11{,}627 & 11{,}625 & $-2$  & $-0.017$ & 492 & 493 & $+1$ \\
    2021 & 15{,}643 & 15{,}641 & $-2$  & $-0.013$ & 666 & 667 & $+1$ \\
    2022 & 17{,}216 & 17{,}223 & $+7$  & $+0.041$ & 700 & 702 & $+2$ \\
    2023 & 19{,}628 & 19{,}625 & $-3$  & $-0.015$ & 719 & 721 & $+2$ \\
    2024 & 22{,}372 & 22{,}371 & $-1$  & $-0.004$ & 823 & 823 & $0$ \\
    \bottomrule
  \end{tabular}
\end{table}

One limit of this check must be stated plainly.
The most recent official report covers 1990--2024 only, so the headline 2025 figures---24{,}458 total and 901 damaging reports---have no published FAA counterpart to be validated against, and the validation above extends no further than 2024.
This compounds the report-lag limitation in Section~\ref{sec:limitations}: 2025 is both the year most exposed to right censoring in our freeze and the one year of the series that no external publication can currently corroborate.
Readers should treat the 2025 values as provisional until the next official multi-year report appears.

\section{Discussion}
\label{sec:discussion}

The aggregates mostly confirm things the wildlife-hazard literature already treats as settled.
What the open tables add is a single, fixed-definition view of those patterns across the full civil record, so the same claims can be checked without re-implementing the Access export.

Start with the time series (Section~\ref{sec:results}).
The \emph{share} of reports coded as damaging falls across the decades while the absolute count of damaging reports rises.
NWSD reporting has long been largely voluntary; completeness has improved over time, and damaging events have always been more likely to be filed than minor ones~\citep{dolbeer2015reporting}.
Better capture of minor encounters is the leading explanation for the falling share.
A fall in underlying risk is not required to produce that pattern---you get the same shape if operators simply log more of the zero-damage events they used to skip.

The family charts tell a different story than a raw species leaderboard (Figure~\ref{fig:family}); the commercial-only cut does the same (Figure~\ref{fig:family-comm}).
Families that dominate identified report volume are not the families with the highest damage rates among those reports.
That tracks the body-mass and flocking differences DeVault et al.\ documented~\citep{devault_et_al_2011}.
When we mapped FAA \texttt{SPECIES\_ID} prefixes into families for the release, the hierarchy mattered in practice: two-character codes have to win over one-character parents, and unidentified rows have to stay out of the family rates, or the ``songbirds everywhere'' picture swallows the deer and waterfowl damage signal.
Volume-ranked species lists are therefore a weak place to aim mitigation budgets if the goal is fewer damaging reports rather than fewer reports of any kind.

Airports show the same volume--damage split in another form.
Across the 452 airports that meet the report threshold, the rank correlation between reported volume and damage rate is negative ($\rho = -0.44$, Section~\ref{sec:results}).
That number is easy to misread as a hazard ranking.
It is not.
Airport totals mix traffic, fleet mix, surrounding habitat, and reporting culture; minor-strike capture is plausibly most complete at high-volume hubs with formal wildlife programs, which pulls damage \emph{rates} down even when damaging \emph{counts} remain large.
The resulting ordering is not a normalized hazard index.
Volume-ranked airport lists overstate hazard at large stations and understate it at smaller ones---exactly the trap that appears when a hub with thorough logging sits next to a quieter field that mainly files the events that hurt.
Land use around the airfield, not position on such a list, is the framing under which airport wildlife management actually operates~\citep{blackwell_et_al_2009}.

A recent systematic review reaches similar conclusions about how fragmented strike evidence still is, and about the value of reproducible, openly documented aggregates~\citep{systematic_review_2025}.
Open tables will not replace the FAA annual narrative, but they do let journalism and secondary research work from the same rollups without redistributing full microdata.

\section{Limitations}
\label{sec:limitations}

\begin{itemize}
  \item Report-based data, not a complete census of wildlife strikes.
  \item Reporting is differential by damage. Under a largely voluntary system, damaging strikes are reported at far higher rates than harmless ones~\citep{dolbeer2015reporting}. Every damage rate in this paper is therefore an upper bound on the corresponding conditional probability of damage given a strike. The bias is largest for taxa whose harmless encounters are least likely to be noticed at all, such as deer and other large mammals, and smallest where operator practice makes reporting close to universal.
  \item Family rates condition on species identification. Identified species supply 209{,}778 of 347{,}575 reports, or 60.4\%; the remaining 39.6\% carry no usable \texttt{SPECIES\_ID}. In aggregate this selection is mild and runs in the unexpected direction: 6.10\% of identified reports are coded damaging against 6.70\% of unidentified ones. It is not uniform across families. Unidentified ``small bird'' reports are overwhelmingly songbird-type and overwhelmingly harmless, so the 41.1\% songbird figure is a share of identified reports only.
  \item Right censoring and report lag. NWSD accrues reports with a lag, so a pull dated 2026-07-04 holds an incomplete 2025. The 2025 series high is a lower bound on the eventual total. The 2025 damage share is unstable if damaging and non-damaging reports arrive at different lags.
  \item Damage coding quality varies; \texttt{M?} is treated as damaging.
  \item No flight-hour or movement denominator---rates are not true operational risk rates.
  \item Aircraft grouping uses free-text \texttt{AIRCRAFT}, not standardized ICAO type designators.
  \item Airport ID ambiguity and unknown-airport rows.
  \item Civil/GA composition and reporting culture change over time.
  \item Website explorers may change; the DOI snapshot is canonical.
  \item Independent pipeline may introduce mapping errors relative to official FAA annual PDFs. Section~\ref{sec:validation} bounds this for annual totals and damaging counts over 2015--2024 against the official 1990--2024 report; no comparable check is possible for 2025 or for the species, airport, and aircraft facets.
\end{itemize}

\section{Conclusions}
\label{sec:conclusions}

We provide open, documented aggregates of U.S.\ FAA wildlife-strike reports and re-measure across the full national record the established finding that species and airports ranked by strike volume are not those ranked by damage: the families with the highest conditional damage rates (deer, waterfowl) contribute a small fraction of reported strike volume, and across 452 airports reported strike volume and damage rate are negatively rank-correlated.
The 2025 reporting year sets series highs in both total and damaging counts under our definitions, but it continues a long-run rise in reporting volume rather than standing apart from it: annual totals run 19{,}625 in 2023, 22{,}371 in 2024, and 24{,}458 in 2025, a 9.3\% year-on-year increase in line with the growth of the record, and no year-level maximum in a series that has risen for three decades should by itself be read as a change in underlying hazard.
Because all rates are per reported strike, exposure-normalized follow-on work (per movement or per flight hour) remains the natural next step; the versioned DOI release is intended to make such reuse straightforward.

\section{Data availability}
\label{sec:availability}

Aggregated tables supporting this article are available under CC~BY~4.0 at Zenodo~\citep{flightfinder_zenodo}:
\begin{quote}
\url{https://doi.org/10.5281/zenodo.21758280}
(concept DOI for latest: \url{https://doi.org/10.5281/zenodo.21347859})
\end{quote}
Primary source: U.S.\ FAA National Wildlife Strike Database (public domain)~\citep{faa_nwsd_portal}.
Interactive exploration (non-archival): \url{https://himaxym.com/safety/wildlife-strikes}.
Narrative companion~\citep{flightfinder_story}: \url{https://himaxym.com/stories/bird-strike-damage-rates}.

\paragraph{Code availability.}
The processing scripts that build the aggregates, figures, and statistics from the public FAA source---aggregation, family taxonomy, freeze CSV export, figure generation, and interval/correlation computation---are released as an open code annex~\citep{code_annex}: \url{https://gist.github.com/disclaimer8/330a1032781a68acb1e55181f05dc77c}.

\paragraph{Snapshot.}
This article uses the FlightFinder aggregate seed stamped \texttt{generatedAt=2026-07-04} (\texttt{wildlife-strikes.json} in the archived release).
The archived Zenodo dataset and the title of this article are both stated as 1990--2026, because the published aggregation covers incident years 1990--2026 and the seed includes a partial 2026.
The title window is therefore the span of the aggregation, not the trend window: all trend statistics in Section~\ref{sec:results} use complete incident years 1990--2025 only.
The cross-sectional facets---airport, species, animal family, aircraft, and phase of flight---are computed over the full seed and cover 1990--2026 (Section~\ref{sec:data}).
Machine-readable freeze tables are included in the archived release (\texttt{freeze-2026-07-04/}).

\section*{Competing interests}
The author develops FlightFinder (\url{https://himaxym.com}), a commercial aviation-safety and flight-search product.
This analysis uses only public FAA data; no airline or airport funded the study.

\section*{Author contributions}
D.K.\ designed the pipeline, produced aggregates, and wrote the manuscript.

\section*{Use of generative AI language tools}
Generative AI language tools---a large language model---were used to draft and revise the text of this manuscript.
The data pipeline, the aggregate definitions, and every figure and number reported here were produced by the author's own code against the frozen dataset of Section~\ref{sec:availability} and checked by the author; they are reproducible from the published aggregates and the open code annex~\citep{code_annex}, and the annual totals and damaging counts are additionally compared against the official FAA report in Section~\ref{sec:validation}.
Such tools are not listed as authors.
The author takes full responsibility for the entire contents of this article, irrespective of how any part of it was generated.

\bibliographystyle{plainnat}
\bibliography{references}

\appendix
\section{Animal family prefix table}
\label{app:families}
\begin{center}
\begin{tabular}{lll}
\toprule
Slug & Label & SPECIES\_ID prefixes \\
\midrule
deer & Deer & 1G \\
bats & Bats & 1C \\
other-mammals & Other mammals & 1 \\
waterfowl & Waterfowl & J \\
raptors & Raptors & K \\
owls & Owls & R \\
gulls & Gulls \& terns & NE \\
shorebirds & Shorebirds & N \\
doves & Pigeons \& doves & O \\
songbirds & Perching songbirds & Y, Z \\
herons & Herons \& egrets & I \\
--- & Other & ELSE \\
\bottomrule
\end{tabular}
\end{center}

\section{Damage level codes}
\label{app:damage}
\begin{center}
\begin{tabular}{lll}
\toprule
Code & Label & Counted damaging? \\
\midrule
N & None & No \\
M & Minor & Yes \\
M? & Uncertain & Yes \\
S & Substantial & Yes \\
D & Destroyed & Yes \\
(null) & --- & No \\
\bottomrule
\end{tabular}
\end{center}

\end{document}